\documentclass[10pt,conference]{IEEEtran}
\IEEEoverridecommandlockouts
\usepackage{cite}
\usepackage{amsmath,amssymb,amsfonts}
\usepackage{graphicx}
\usepackage{textcomp}
\usepackage{xcolor}
\usepackage[utf8]{inputenc}
\usepackage[T1]{fontenc}
\usepackage{url}
\usepackage{booktabs}
\usepackage{amsfonts}
\usepackage{nicefrac}
\usepackage{microtype}
\usepackage{xcolor}
\usepackage{amssymb}
\usepackage{multirow}
\usepackage{amsmath}
\usepackage{graphicx}
\usepackage{comment}
\usepackage{array}
\usepackage{colortbl}
\usepackage{subcaption}
\usepackage{caption}
\usepackage{enumitem}
\usepackage{pifont}
\usepackage[most]{tcolorbox}
\usepackage{wrapfig}
\usepackage{float}
\usepackage{enumitem}
\usepackage{fontawesome}
\usepackage{tikz}
\usepackage{listings}
\usepackage[hidelinks,breaklinks=true]{hyperref}
\usepackage{xurl}      %

\def\BibTeX{{\rm B\kern-.05em{\sc i\kern-.025em b}\kern-.08em
    T\kern-.1667em\lower.7ex\hbox{E}\kern-.125emX}}
\begin{document}

\title{Better Harnesses, Smaller Models: Building 90\% Cheaper Agents via Automated Harness Adaptation
}

\author{
\IEEEauthorblockN{
Chenyang Yang,
Xinran Zhao, 
Tongshuang Wu, and
Christian K\"astner}
\IEEEauthorblockA{\textit{Carnegie Mellon University}}
}

\maketitle

\newcommand{\toolname}[0]{\textsc{Morphic}}
\newcommand\circleone{\ding{192}}
\newcommand\circletwo{\ding{193}}
\newcommand\circlethree{\ding{194}}
\newcommand\circlefour{\ding{195}}
\newcommand\circlefive{\ding{196}}
\newcommand\circlesix{\ding{197}}
\newcommand\circleseven{\ding{198}}
\newcommand\circleeight{\ding{199}}

\definecolor{mybgcolor}{HTML}{E6F2F0}    %
\definecolor{myframecolor}{HTML}{FF8552} %
\definecolor{mytitlecolor}{HTML}{2C7873} %

\definecolor{tblheader}{HTML}{F5E6DC}   %
\definecolor{tblheadtext}{HTML}{5B3A1E}  %
\definecolor{adaptgray}{HTML}{999999}    %

\newcommand{\nbc}[3]{
 {\colorbox{#3}{\bfseries\sffamily\scriptsize\textcolor{white}{#1}}}
 {\textcolor{#3}{\sf\footnotesize$\blacktriangleright$\textit{#2}$\blacktriangleleft$}}
 }

\newcommand\todo[1]{\nbc{TODO}{#1}{red}}
\newcommand{\cyang}[1]{\nbc{CY}{#1}{teal}}
\newcommand{\swcomment}[1]{\nbc{SW}{#1}{blue}}
\newcommand{\ck}[1]{\nbc{CK}{#1}{orange2}}
\newcommand{\mxl}[1]{\nbc{MS}{#1}{violet}}
\newcommand{\cma}[1]{\nbc{CM}{#1}{orange}}

\newcommand{\delete}[1]{\textcolor{deeporange}{\st{#1}}}
\newcommand{\new}[1]{{#1}}

\newcounter{hyp}
\newcommand{\Hypothesis}[1]{%
  \refstepcounter{hyp}%
  \textbf{Hypothesis~\thehyp}: {#1}\label{hyp:\thehyp}%
}

\newcounter{rq}
\newcommand{\RQ}[1]{%
  \refstepcounter{rq}%
  \textbf{RQ~\therq}: {#1}\label{rq:\therq}%
}

\definecolor{tagbg}{RGB}{245,237,255}
\definecolor{tagtext}{RGB}{110,72,170}

\DeclareRobustCommand{\dimtag}[1]{%
  \tcbox[
    on line,
    nobeforeafter,
    tcbox raise base,
    colback=tagbg,
    colframe=tagbg,
    arc=3pt,
    boxsep=0pt,
    left=3pt,
    right=3pt,
    top=1pt,
    bottom=1pt,
    boxrule=0pt,
    fontupper=\sffamily\bfseries\footnotesize\color{tagtext},
    before upper={\strut}
  ]{#1}%
}

\DeclareRobustCommand{\tabledimtag}[1]{%
  \tcbox[
    on line,
    nobeforeafter,
    tcbox raise base,
    colback=tagbg,
    colframe=tagbg,
    arc=2pt,
    boxsep=0pt,
    left=2pt,
    right=2pt,
    top=0pt,
    bottom=0pt,
    boxrule=0pt,
    fontupper=\sffamily\bfseries\scriptsize\color{tagtext}
  ]{#1}%
}

\lstset{
  language=Python,
  keywords={},
  commentstyle=\textbf,
  stringstyle=\text,
  extendedchars=false,
  basicstyle=\ttfamily,
  columns=fullflexible,
  frame=single,
  breaklines=true,
  breakatwhitespace=true,
  breakindent=0\dimen0,
  literate={'}{{\textquotesingle}}1 {`}{{\textasciigrave}}1 {"}{{\textquotedbl}}1,
}

\begin{abstract}
Frontier LLM agents are automating many business tasks, but their high inference cost makes large-scale deployment unsustainable.
Small language models (SLMs) offer a cheaper alternative, yet they typically fall short when swapped into a harness designed for a frontier LLM.
We show that for many routine business tasks, SLM agents can match LLM performance at 90\% lower cost,
when paired with an adapted harness that can be automatically discovered by a meta agent.
The key insight is that much of the task difficulty is shared across instances and can be lifted from the model into the harness via tailored instructions, tools, and orchestration loops.
To study this systematically, we create a framework that maps agent failure modes to harness adaptation strategies, and build a harness optimizer that automatically discovers effective adaptations from failure trajectories.
Across seven business-oriented agentic tasks and three SLM families, we found optimized harnesses significantly improve performance on 16 of 21 task-SLM pairs, with seven pairs closing the SLM-LLM performance gap and the best SLM agent recovering 89.7\% of LLM performance at 4\% of the cost.
\footnote{Code available at \faGithub~\url{https://github.com/malusamayo/migration-analysis}.}
Our analysis further shows that adaptation works best for tasks with more repetitive workflows and for SLMs with sufficient base capabilities.
Together, these results suggest that harness adaptation can expand the practical deployment range of SLM agents in routine business tasks.
\end{abstract}


\section{Introduction}
\label{sec:intro}

\begin{figure*}[t]
  \centering
  \includegraphics[width=0.95\linewidth]{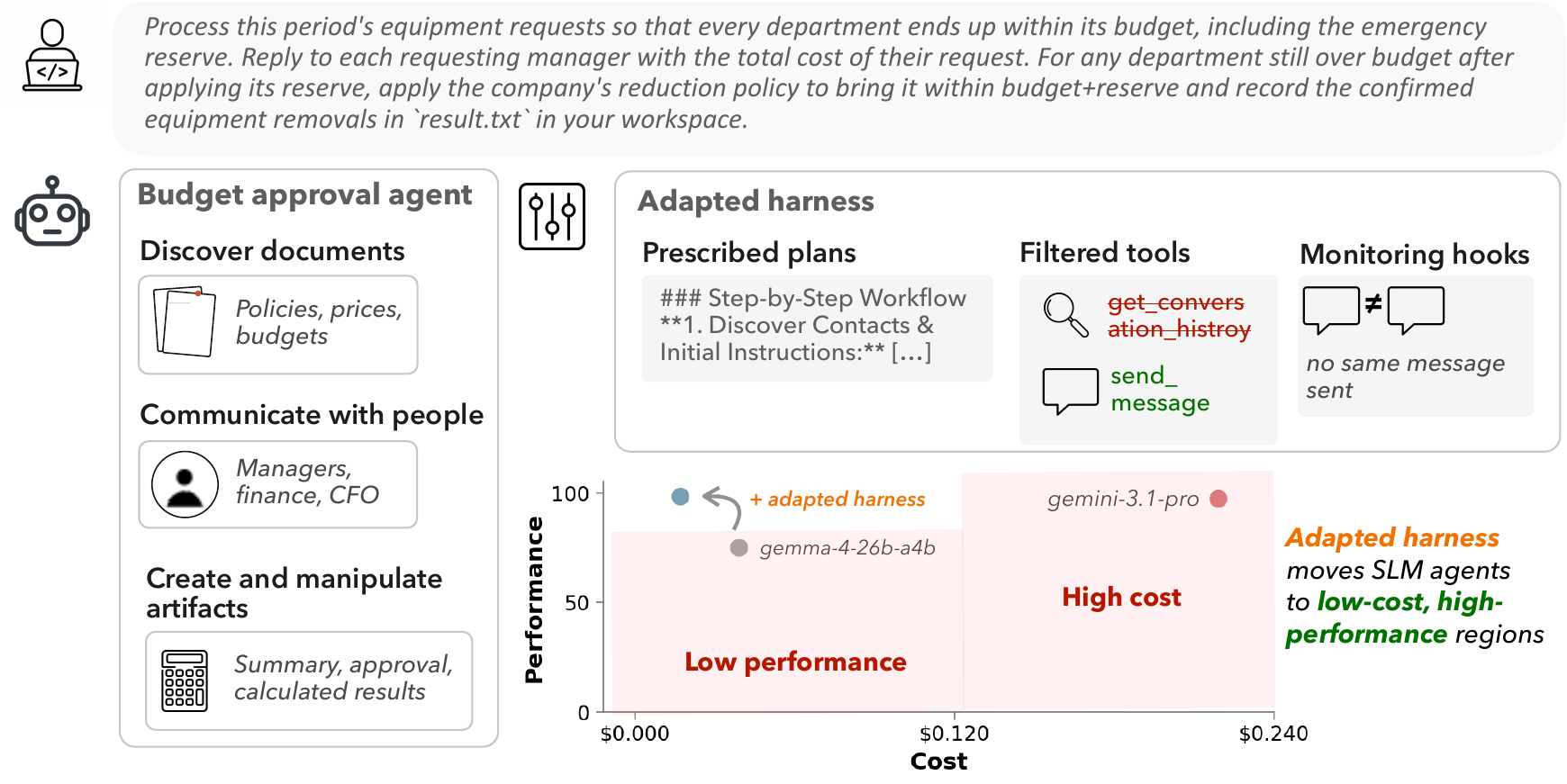}
  \caption{%
    Small language models often perform poorly when naively swapped into an agent harness designed for frontier LLMs. 
    An adapted harness can provide detailed plans to guide workflow execution, a reduced tool set to avoid invocation mistakes, and monitoring hooks to avoid loops.
    These adaptations make an SLM on par with an LLM with 8\% of the cost. 
  }
  \label{fig:example}
  \vspace{-10pt}
\end{figure*}

Large language models (LLMs) power many agentic systems that are increasingly deployed in real-world business workflows~\cite{aws_cox_auto_agentcore_2026, salesforce_engine_agentforce}.
While frontier models deliver impressive performance, they incur substantial inference cost, high latency, and data-privacy concerns --
businesses have poured billions of dollars into incorporating agents into their workflows~\cite{reuters2026anthropic380b},
and that continuous spending can quickly become unsustainable~\cite{bratton2026uberclaudecodebudgets}.

Small language models (SLMs)\footnote{We use SLMs operationally for cheaper, smaller, open-weight models that can be deployed locally. Our experiments use models with 3B to 8B (8B to 30B) active (total) parameters (e.g., \texttt{qwen3-30b-a3b}~\cite{yang2025qwen3}). In comparison, frontier open-sourced models have more than 1T model parameters~\cite{team2026kimi, deepseekai2026deepseekv4}.}
have emerged as an alternative for replacing LLMs on agentic tasks, with their lower cost, better latency, and less privacy concerns~\cite{belcak2025small}.
While SLMs fall short of LLMs on general-purpose agentic coding tasks~\cite{openhands_index_2026}, 
many business-relevant tasks do not require open-ended creativity, but only reliable execution of routine workflows within a constrained environment~\cite{belcak2025small} -- this makes SLMs an attractive option for replacing LLMs in these contexts.

As a concrete example, suppose we want to deploy an agent to collect, review, and communicate various budget requests in a company, a routine business function that many try to automate.
We can deploy an agent with a frontier LLM that will perform very well (97.3\% accuracy with \texttt{gemini-3.1-pro} on curated evaluation data), but is also expensive (\$0.22 per query) to deploy in production.
Alternatively, we can deploy the same agent with an SLM.
The agent, however, performs worse (75.0\% with \texttt{gemma-4-26b-a4b}) and cannot make a reliable replacement (Figure~\ref{fig:example}).

In this paper, we demonstrate that, for many routine agentic tasks in business settings,
\textbf{we can build specialized SLM agents that are 90\% cheaper yet with on-par performance with LLM agents}.
This is achieved by pairing SLMs with \textit{well-adapted harnesses} that can be discovered automatically with a meta agent.
This is because, for these tasks, a lot of task difficulty is shared across instances and can be lifted from the model to the harness, through (automatically) tailored instructions, tools, and orchestration loops.
In our budget-approval example, while SLMs may struggle to create complex plans or select appropriate tools on their own, 
a well-designed harness can scaffold an SLM with a clear plan skeleton, a reduced tool set, and enforced constraints via hooks, improving SLM performance to 98.3\% (Figure~\ref{fig:example}, right).

To build effective specialized agents with SLMs, we first need to understand what makes a task hard and how that difficulty can be absorbed into the harness. 
To this end, we synthesize a list of agent failure modes and harness adaptation strategies from existing literature (Section~\ref{sec:theory}),
covering common failure modes such as \dimtag{tool-use} and \dimtag{knowledge} failures,
and adaptation strategies on contexts, tools, and orchestration loops.
This gives us a framework for diagnosing why an SLM fails on a task and identifying and explaining which harness adaptations are likely to help:
For example, when we observe an SLM struggle with choosing tools from a large set (\dimtag{tool-use} failures), 
we can adapt the harness by creating high-level tools and filtering unused ones.

While harness adaptations can be effective for many tasks, it is unclear when harness adaptation is sufficient to close the SLM-LLM gap, and which task or model conditions make adaptation more effective.
To systematically understand if, when, and how we can build (cost-)effective SLM agents with harness adaptation, 
we conduct an empirical study across seven curated business tasks and three SLMs from different model families (Section~\ref{sec:experiments}).
Rather than engineering a harness by hand for each task–SLM pair, we develop a harness optimizer that automatically discovers effective adaptations:
We design the harness optimizer as a meta-agent that systematically diagnoses agent failures and proposes targeted harness adaptations (Section~\ref{sec:method}), 
building on recent advances of automated harness optimization~\cite{zhang2025darwin, lee2026meta}.
This allows us to systematically study the effectiveness of SLM harness adaptation without laborious manual engineering.
From the study, we find that,
\begin{itemize}[leftmargin=*]
   \item \textbf{Harness adaptation substantially improves the cost--performance tradeoff of SLM agents.}
    Optimized harnesses significantly improve performance on 16 out of 21 SLM-task pairs, with seven pairs closing the SLM-LLM performance gap. 
    The best adapted SLM can recover 89\% of LLM-agent performance with 96\% cost reduction.

   \item \textbf{Harness adaptation is grounded in common failure modes.}
   Successful adaptations most often address \dimtag{instruction-following} (81\%) and \dimtag{knowledge} (81\%) failures. 
   The dominant strategies include adding contexts (86\%), creating new tools (43\%), and managing tools (29\%).

   \item \textbf{Harness adaptation is more effective for less diverse tasks.}
    Optimized harnesses work best when task instances share a stable workflow (+21.1\% accuracy when we go from the most diverse task setting to the least diverse setting). 
    
   \item \textbf{Harness adaptation is more effective for SLMs with stronger base capabilities.}
    More capable SLMs benefit more (+48.8\% vs. +15.5\%) from harness adaptation, suggesting that harnesses can absorb substantial task difficulty but cannot fully replace missing core model capabilities.
\end{itemize}

In summary, this paper contributes:
\begin{itemize}[noitemsep,topsep=0pt, leftmargin=*]
  \item A task suite of 7 diverse agentic tasks designed for studying agent deployment on repetitive business tasks.
  \item Empirical findings from experiments on 7 tasks × 3 SLMs, showing that harness adaptations recover substantial LLM performance at a fraction of the cost, and that we can characterize when and why adaptations work.
  \item A framework mapping failure modes to harness adaptation strategies, explaining how the adapted harnesses are effective. 
  \item A harness optimizer that automatically searches for effective harness adaptations for SLM agents. 
\end{itemize}

\section{Characterizing Language Model Agent:\\ Failures and Adaptations}
\label{sec:theory}

To identify promising harness adaptations, we first need to understand why agents fail and what harness adaptations are available. 
In this section, we first establish a framework to characterize agent failures and map them to relevant harness adaptation strategies.

\subsection{Agent failure modes}
\label{sec:capabilities}

\looseness=-1
Failures in agentic task outcomes come from one or multiple \textit{local} failures in the trajectory. 
For example, a budget-approval agent that approves invalid requests may have selected the wrong tool at one step, violated a policy instruction at another, relied on an incorrect assumption, lost track of an earlier budget limit near the end, or created an infeasible plan in the beginning.

We organize such local failure modes by the \textit{capabilities} they implicate, such as \dimtag{tool-use}, \dimtag{knowledge}, \dimtag{instruction-following}, \dimtag{long-context}, and \dimtag{planning}.
We derive these capabilities by surveying recent model cards~\cite{openai2025gpt52, anthropic2026opus47, qwen35blog} and the agentic benchmarks they report (e.g., \cite{patil2025bfcl, bandi2026mcp, zhou2023instruction, wei2024measuring, jain2024livecodebench, patwardhan2025gdpval, merrill2026terminal, zhou2023webarena, wei2025browsecomp}). 
Indexing failures by capability allows us to link local failures to the underlying model abilities that may have broken down, and to the adaptations likely to address them.
We next describe each capability-indexed failure mode in detail (Figure~\ref{fig:taxonomy}, left):

\begin{figure*}[ht]
    \centering
    \includegraphics[width=0.85\textwidth]{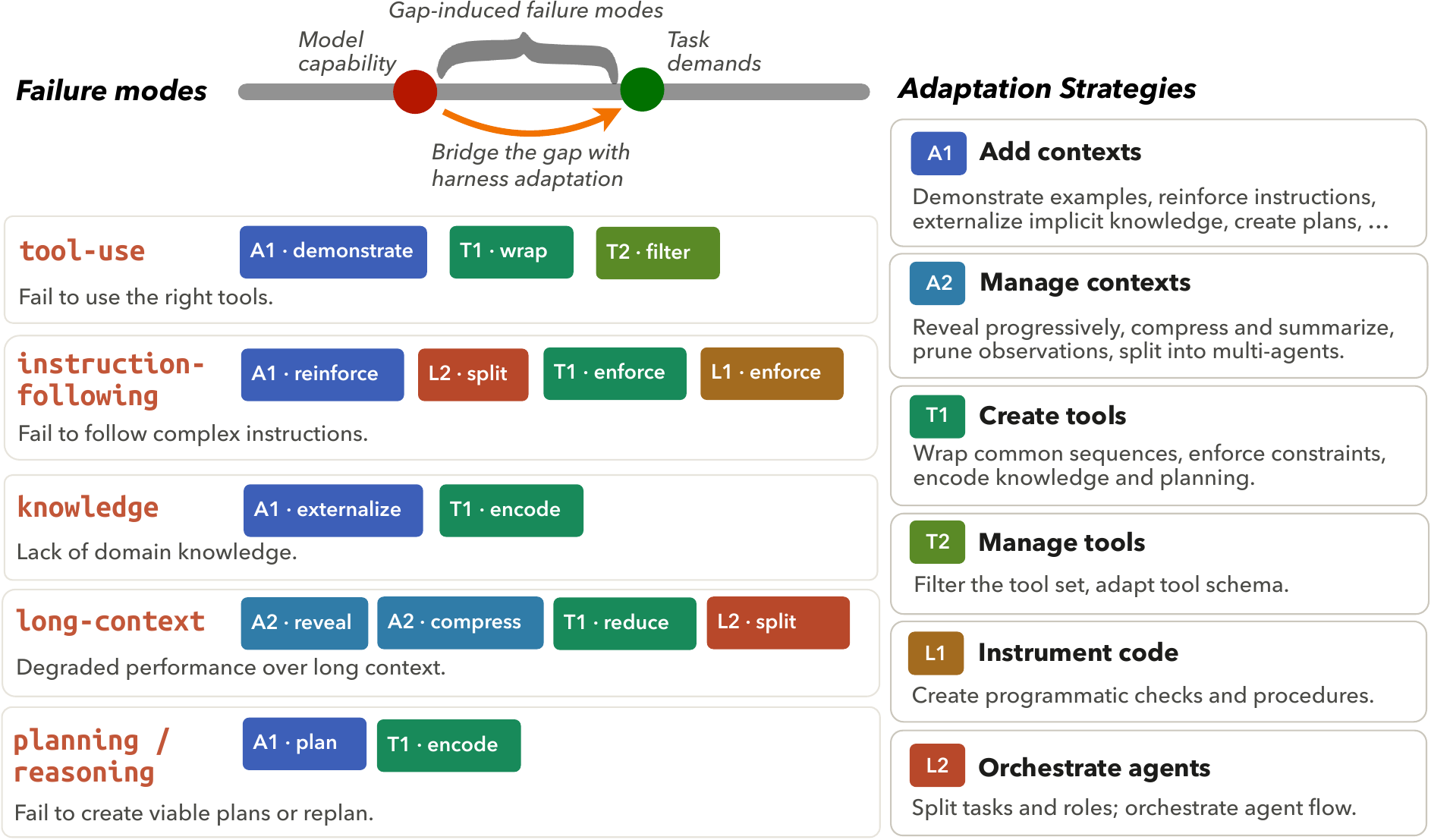}
    \caption{\textbf{Mapping Agent Failure Modes to Harness Adaptations.} When model capabilities fall short of task demands, this mismatch manifests as agent failures in trajectories. These gaps can be bridged with harness adaptation, which we group into three categories: adapting contexts, tools, and agent loops. 
    }
    \vspace{-15pt}
    \label{fig:taxonomy}
\end{figure*}

\paragraph{\dimtag{tool-use} failures}
Tool-use failures occur when the model selects the wrong tool, produces a malformed tool call, or uses a tool incorrectly.
They are especially common when the tool set is large, tool schemas are complex, or the task requires composing multiple tools.

\paragraph{\dimtag{instruction-following} failures}
Instruction-following failures occur when the model violates provided system or user instructions, which can include task requirements, output formatting rules, workflow guidelines, etc.
They are especially common when instructions are long and complex.

\paragraph{\dimtag{knowledge} failures}
Knowledge failures occur when the model lacks the domain, environment, or commonsense assumptions needed to complete an under-specified task.
Because real-world instructions are often under-specified~\cite{yang2025prompts}, agents are expected to fill in missing details from pretraining knowledge or from interacting with the environment.
When this knowledge is missing, agents may call the wrong tools, misinterpret task constraints, or follow sub-optimal workflows.

\paragraph{\dimtag{long-context} failures}
Long-context failures occur when the model fails to retrieve and use information from a long trajectory.
Failures may appear as forgetting earlier instructions, repeating already-failed actions, overlooking relevant observations, or being distracted by irrelevant context.

\paragraph{\dimtag{planning / reasoning} failures}
Planning and reasoning failures occur when the model cannot decompose a high-level goal into appropriate sub-goals, choose an effective next action, or revise its plan after new observations.
These failures are central to agentic tasks because the model is expected to repeatedly decide what to do next based on partial and evolving information.
They may appear as premature answers, inefficient exploration, failure to backtrack, or local actions that do not contribute to the overall goal.

\textbf{Comparison with existing failure analysis.}
Prior work has also proposed empirical taxonomies for understanding agent failures.
Cemri et al.~\cite{cemri2025multi} introduce MAST, a taxonomy of multi-agent failures covering system design issues, inter-agent misalignment, and task verification.
Zhu et al.~\cite{zhu2025llm} propose AgentErrorTaxonomy, which attributes single-agent failures to memory, reflection, planning, action, and system-level modules.
In contrast, we index failures by model \textit{capabilities}, making it easier to connect observed failures to capability gaps and targeted harness adaptations.

\subsection{Agent harness adaptations}

To address observed agent failures, offloading the task demand from the model to the harness can be an effective strategy.
For example, a \dimtag{knowledge} failure may be mitigated by \textit{adding contexts} of domain instructions, a \dimtag{tool-use} failure by transforming the available tools, a \dimtag{long-context} failure by hiding irrelevant observations, and an \dimtag{instruction-following} failure by enforcing constraints outside of the model outputs.

We present an overview of harness adaptation strategies that recur in existing literature and popular harnesses (e.g.,~\cite{anthropic_claude_code_overview_2026,openai_codex_docs_2026}).
We group these strategies by the harness component they modify: 
Context adaptations change what information is presented to the model, tool adaptations change the actions available to the model, and agent-loop adaptations change the control logic around model inference (Figure~\ref{fig:taxonomy}, right).

\paragraph{Context adaptations}
Context adaptations shape the context a model acts on, including system prompts, user prompts, tool descriptions, memory, and other observable information like files and databases in the environment.

\looseness=-1
\textbf{Adding contexts} is the most straightforward and commonly used adaptation strategy: 
It helps with \dimtag{knowledge} gaps by externalizing implicit knowledge into instructions, 
with \dimtag{planning} gaps by providing explicit plans,
with \dimtag{tool-use} gaps by providing more detailed tool descriptions and examples,
and with \dimtag{instruction-following} gaps by reinforcing key constraints.
It has been well studied in the literature as prompt engineering and context engineering~\cite{zhang2025agentic}, and widely used by practitioners to steer agent behaviors (e.g., AGENTS.md~\cite{gloaguen2026evaluating}).
Adding contexts, however, also incurs \dimtag{long-context} burden and more complex \dimtag{instruction-following}.
\textbf{Managing contexts} properly is thus essential to prevent overwhelming SLMs.
Common strategies include revealing contexts progressively~\cite{anthropic_skill_authoring_best_practices_2026}, offloading contexts into external storage~\cite{zhang2025recursive}, compressing contexts with summaries~\cite{kang2025acon}, pruning irrelevant observations~\cite{xiao2025improving}, and separating contexts into distinct components like multi agents~\cite{xu2025boad}.

\paragraph{Tool adaptations}
Tool adaptations reshape the action space an agent operates in -- they are commonly implemented as (1) MCP servers~\cite{anthropic_model_context_protocol_2024}, (2) command-line tools, or (3) executable scripts (e.g., as in agent skills~\cite{anthropic_skill_authoring_best_practices_2026}).
\textbf{Creating tools} is a powerful strategy to mitigate \dimtag{tool-use} gaps: 
By wrapping recurring complex tool use patterns and sequences into simpler interfaces, we can support easier and more stable tool use~\cite{wang2025inducing}.
It also helps with \dimtag{knowledge} and \dimtag{planning} gaps by encoding domain knowledge and plans into the tool implementation,
with \dimtag{instruction-following} gaps by enforcing certain constraints programmatically, 
and with \dimtag{long-context} gaps by providing more compact tool observations.
When many tools are created, we will also need to \textbf{manage tools} properly.
By filtering out irrelevant tools, we can make tool selection easier and more efficient~\cite{yang2024agentoccam}.
By tailoring tool schema, we can make tools more intuitive and easier to use~\cite{lee2025don}.

\paragraph{Agent loop adaptations}
Finally, we can also adapt the agent loop structure to mitigate failures from having a single unreliable model to take all actions with accumulating contexts.
\textbf{Instrumenting code} is a common strategy to mitigate the unreliability of the plain agent loop. 
This can be done by adding deterministic checks (e.g., agent hooks~\cite{anthropic_claude_code_hooks_2026}) to the loop that can be triggered when certain conditions are met, such as before or after calling certain tools.
This helps with briding \dimtag{instruction-following} gaps by enforcing some constraints programmatically.
In safety contexts, this is also referred to as symbolic guardrails~\cite{hong2026symbolic} to catch unsafe behaviors.
\textbf{Orchestrating agents} is another strategy to go beyond single agent limitations,
through building multi-agent systems~\cite{kim2025towards}, often to help with \dimtag{long-context} problems.
Many different topologies and orchestration strategies exist in the literature:
For example, we can build systems where there is one main agent that spawns and orchestrates subagents~\cite{fourney2024magentic}, 
systems where different agents are peers and can communicate~\cite{hong2023metagpt},
or systems with independent agents that communicate in the end.
Multi-agent system is an active research field -- a more comprehensive discussion can be found in~\cite{kim2025towards}.

\textbf{Trade-offs of harness adaptations.}
Creating more complex harnesses comes with costs: 
For example, while \textbf{adding contexts} is straightforward and often effective, it can lead to new failures on \dimtag{long-context} and \dimtag{instruction-following} due to growing context complexity.
\textbf{Managing contexts} can help alleviate growing contexts, but comes with risks of additional \dimtag{planning / reasoning} failures as models are tasked with more exploration work.
\textbf{Creating tools} provides more powerful scaffolds, but can exacerbate \dimtag{tool-use} failures when many tools are created. 
\textbf{Orchestrating agents} may increase inference cost and latency, and create pressure on \dimtag{tool-use} for managing agents.
Indeed, there is no one-size-fits-all solution when we create harnesses for SLMs. 
Instead, we need to create harnesses specialized to the task contexts and observed failures, so that we can make deliberate tradeoffs among the complexities introduced for each specific task.

\subsection{Adapting harnesses for small language models}

Our framework is \textit{intentionally} not tied to model size and applies to language model agents broadly.
This is because SLMs are not a static concept: their capabilities continue to improve over time. 
It is therefore more useful to reason about models in terms of \textit{where their capabilities fall on a spectrum} (Figure~\ref{fig:taxonomy}, top) than to assume a static set of SLM limitations that may quickly become outdated.

That said, harness adaptation is especially relevant for small language models. 
This is because adaptation matters most when a task's capability demands exceed what a model can reliably provide, and this mismatch is more common for SLMs.
As we will show in our experiments (Section~\ref{sec:experiments}), 
frontier LLMs are often strong enough that they can perform well with general-purpose harnesses. 
SLMs, by contrast, more often exhibit capability gaps that make creating specialized harness adaptation especially useful.

\section{Automating SLM Harness Adaptation}
\label{sec:method}
The framework in Section~\ref{sec:theory} suggests that many SLM failures can be mitigated through appropriate harness adaptations. 
In practice, however, finding the right adaptation is challenging because the design space is large.
Identifying the most effective strategy requires extensive trial and error, 
as it is common to have multiple failure modes intertwined (e.g., failing to use appropriate tools can interact with growing contexts), and there are usually multiple plausible adaptations that might address the observed issues (Figure~\ref{fig:taxonomy}). 
This process can be time-consuming for human engineers, especially when the harness may need to be updated repeatedly with evolving models~\cite{anthropic_harness_design_long_running_apps_2026}.

\begin{figure}[t]
    \centering
    \includegraphics[width=\linewidth]{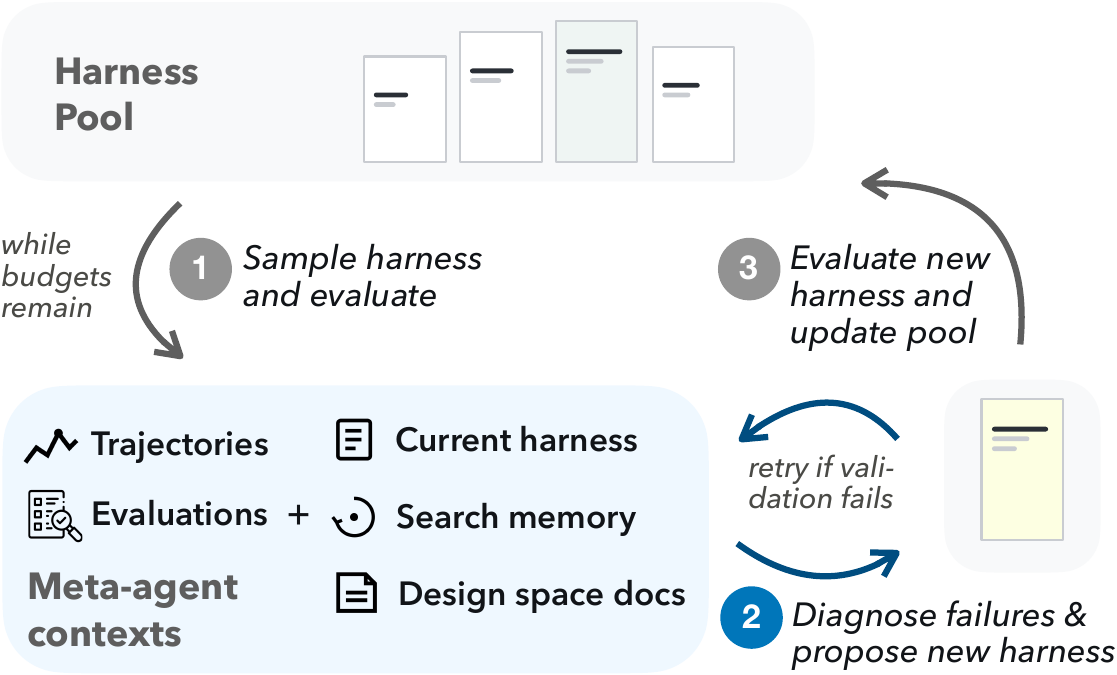}
    \caption{Overview of the automated harness optimization loop.}
    \label{fig:algorithm}
\end{figure}

Learning from the bitter lesson that human-knowledge-based methods ultimately fall short of general methods that leverage search and learning~\cite{sutton2019bitter}, we argue that \textbf{harness design should also be automated as a search problem} driven by data and evaluation, rather than relying on manual trial and error. 
We build on recent advances of automated harness optimization~\cite{zhang2025darwin, lee2026meta, xia2025live}
and create a meta-agent powered harness optimizer that automates the search for identifying effective agent harnesses (Figure~\ref{fig:algorithm}).
The optimizer takes an SLM, a target task with training and validation data, and an initial agent harness as input, and outputs an optimized agent harness.
We expose a general yet concrete search space to the optimizer, and create an optimization loop that searches this space -- we present details of our design below.

\subsection{Search Space}
We instantiate the search space with the API interfaces of \texttt{software-agent-sdk}, an open-source framework for building software agents~\cite{wang2025openhands}.
This gives the meta-agent a concrete set of editable components instead of an unconstrained space.
In our implementation, the harness can modify:
\begin{itemize}[leftmargin=*]
  \item \textbf{Contexts:} system prompts (task instructions, examples), skills, and dynamic contexts that are appended on need.
  \item \textbf{Tools:} primitive tools (file editor, bash, browser), custom tools defined via classes, custom scripts.
  \item \textbf{Hooks:} custom scripts that are triggered based on tool use.
  \item \textbf{Context management:} external file systems, context condensers.
  \item \textbf{Sub-agents:} custom agents with dedicated contexts.
\end{itemize}

These components instantiate a practical subset of the adaptation strategies discussed in Section~\ref{sec:theory}, which makes the search space expressive enough to cover the failures we care about while still being structured enough for reliable editing.

\begin{figure}[t]
    \centering
    \includegraphics[width=\linewidth]{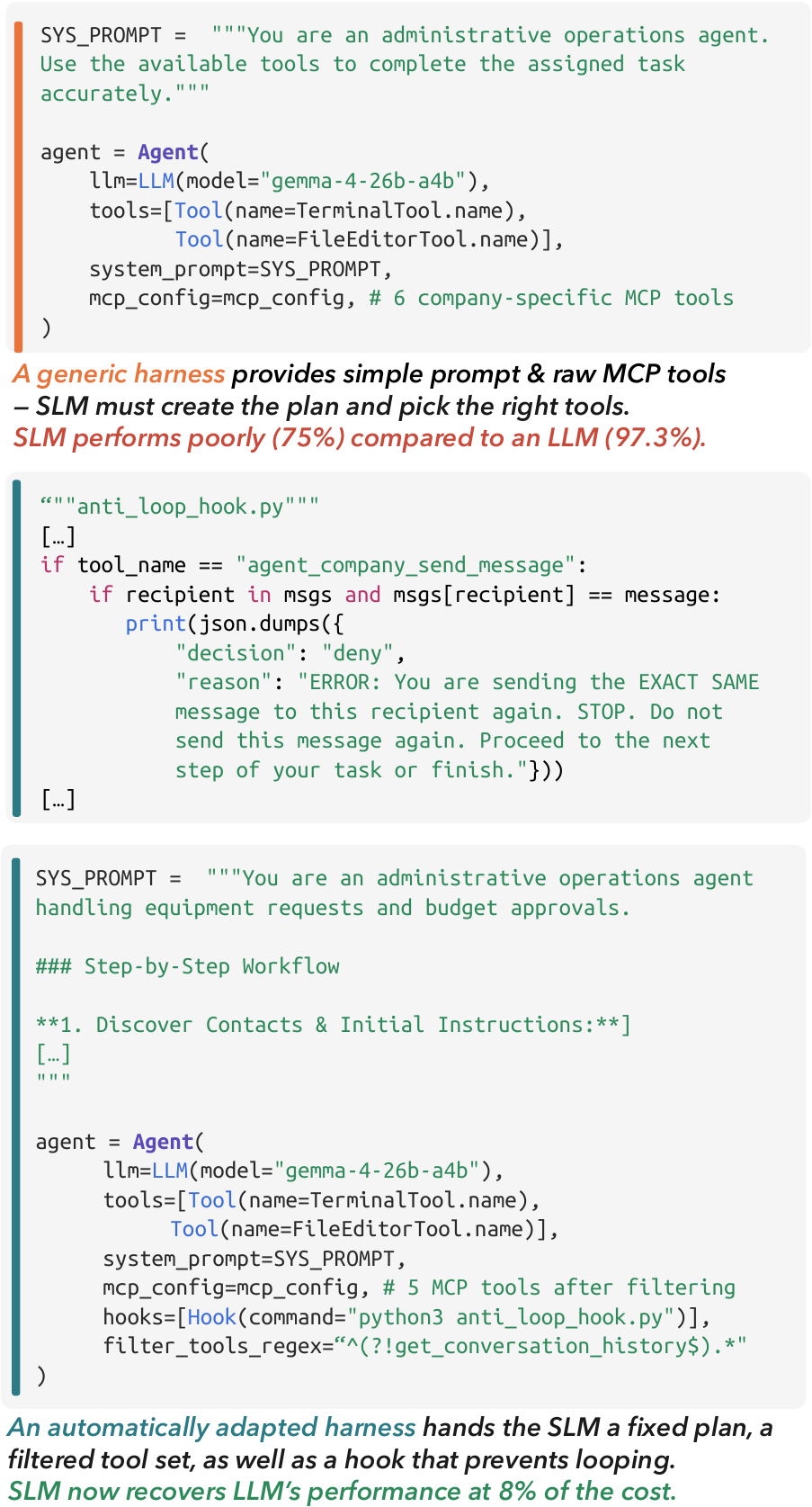}
    \caption{Simplified harness code identified by the meta-agent for the budget-approval task. 
    The harness creates a new Python script as a tool-use hook, filters down tool sets, and creates more detailed workflow instructions.}
    \label{fig:code}
\end{figure}

\subsection{Optimization Loop}
We design an optimization loop that repeatedly (1) samples a harness and evaluates it, (2) diagnoses failures and proposes new harnesses, and (3) evaluates the harnesses (Figure~\ref{fig:algorithm}):
\begin{enumerate}
  \item \textbf{Sample and evaluate a harness.} We maintain a pool of previously tried harnesses and their validation scores. 
  At each iteration, the outer loop selects one harness from this pool for further refinement. 
  We currently use a GEPA-style genetic search procedure~\cite{agrawal2025gepa} to sample candidates from the Pareto front -- i.e., the sampled harness will be optimal on at least one training instance.
  The selected harness is then executed and evaluated on a sampled batch of training instances. 
  We log full trajectories, including tool calls, intermediate observations, outputs. We also save evaluation scores and feedback.
  \item \textbf{Diagnose failures and propose new a harness.} 
  The inner-loop meta-agent then inspects these trajectories together with the sampled harness code and identifies what went wrong. 
  The meta-agent edits the harness by changing harness components, such as adding a tool, inserting a reusable skill, etc.
  The meta-agent receives four main sources of context to make decisions:
    \begin{itemize}[leftmargin=*]
      \item \textbf{Task trajectories:} the latest batch of task trajectories annotated with outcome signals.
      \item \textbf{Current harness:} the full Python implementation of the harness being edited.
      \item \textbf{Search memory:} summaries of past proposals and their observed effects, so the agent does not repeatedly try the same ineffective fix.
      \item \textbf{Design-space documentation:} API references that document the available harness components.
    \end{itemize}
  Before evaluation, we first run a cheap sanity check on the proposed harness to catch invalid code and obviously broken harness implementations. 
  If the proposal fails this check, the meta-agent is asked to repair it and try again up to a maximum number of attempts.
  \item \textbf{Validate and save.} 
  Once the new harness passes the sanity check, we evaluate it on the same sampled batch of training instances and observe if there is any improvement.
  If yes, we run a full evaluation on the validation set and add it to the pool if it improves over prior candidates.
\end{enumerate}

\subsection{Example of an Optimized Harness}
In the budget-approval task (Figure~\ref{fig:example}), the default harness exposed a broad set of MCP tools and requires the model to plan the task end-to-end (Figure~\ref{fig:code}, top). 
This setup works for a frontier LLM but not an SLM that struggles to infer the correct procedure reliably and invoke the correct tools.

\looseness=-1
Our harness optimizer identified a few changes that successfully address the failure modes after 23 iterations. 
It narrowed the action space, rewrites the system prompt into an explicit step-by-step procedure, and introduces a custom \texttt{anti\_loop\_hook.py} safeguard that prevents the model from getting stuck in a loop (Figure~\ref{fig:code}, bottom).
Rather than relying on the SLM to plan and recover from tool-use mistakes on its own, the harness externalizes the fragile parts of the workflow into prompts, filters, and runtime checks. 
The SLM now recovers the frontier model's performance with 8\% of the cost.

\section{Experiments}
\label{sec:experiments}

\begin{table*}[ht]
  \centering
  \caption{We curate a diverse set of seven business-relevant agentic tasks. Each task is grounded in existing benchmarks with verifiable outcomes.}
  \label{tab:task-overview}
  \begin{tabular}{@{}>{\raggedright\arraybackslash}p{1.3cm}>{\raggedright\arraybackslash}p{3.5cm}>{\raggedright\arraybackslash}p{0.5cm}>{\raggedright\arraybackslash}p{4cm}>
  {\raggedright\arraybackslash}p{3cm}>{\raggedright\arraybackslash}p{3.5cm}@{}}
    \toprule
    \textbf{Task} & \textbf{Description}  & \textbf{N} & \textbf{Common workflow} & \textbf{Evaluation criteria} & \textbf{Task creation process} \\
    \midrule
    attendance-auditing
      & Audit attendance and payroll records, then produce a summary spreadsheet.
      & 100
      & Gather records; process data; write summary.
      & Check against ground-truth
      & Create task templates based on TheAgentCompany benchmark~\cite{xu2024theagentcompanybenchmarkingllmagents} and generate new instances. \\
    \addlinespace

    budget-approval
      & Review budget requests against predefined budget constraints.
      & 100
      & Collect requests; retrieve prices and budget policies; reply to requests.
      & Check against ground-truth
      & Create task templates based on TheAgentCompany benchmark~\cite{xu2024theagentcompanybenchmarkingllmagents} and generate new instances. \\
    \addlinespace

    stock-alert
      & Monitor product stock levels and alert under pre-defined conditions.
      & 100
      & Inspect data; compute indicators; send alert.
      & Check against ground-truth
      & Select from LOCA-Bench~\cite{zeng2026loca} and reuse its pipeline to generate new instances.  \\
    \addlinespace

    anomaly-detection
      & Analyze IoT time-series data to identify anomalies.
      & 100
      & Inspect data; run analysis; verify anomalies; upload summary.
      & Check against ground-truth
      & Select from LOCA-Bench~\cite{zeng2026loca} and reuse its pipeline to generate new instances. \\
    \addlinespace

    playwright-testing
      & Generate end-to-end playwright tests for a static website.
      & 100
      & Inspect website; write tests; run and debug failures.
      & Run generated tests
      & Reuse instances from WebGenBench~\cite{lu2025webgenbenchevaluatingllmsgenerating} with complete HTML websites generated by a frontier LLM (\texttt{gemini-3.1-pro}). \\
    \addlinespace

    website-management
      & Complete shopping-site administration tasks through a browser interface.
      & 50
      & Inspect website (in a loop); return exact final answer.
      & Check against ground-truth
      & Sample instances from CMS subset of WebArena~\cite{zhou2023webarena} and keep the instances that a frontier LLM can solve. \\
    \addlinespace

    code-refactoring
      & Refactor a repository based on user instructions.
      & 100
      & Understand repository; inspect relevant files; edit code; iterate.
      & Run tests to check refactored code's AST
      & Reuse instances from RefactorBench~\cite{gautam2025refactorbench}. \\
    \bottomrule
  \end{tabular}%
  \vspace{-10pt}
\end{table*}

We next employ the harness optimizer to study if, when, and how we can build (cost-)effective SLM agents with harness adaptations:

\begin{itemize}[leftmargin=*]
    \item \textbf{RQ1: To what extent can harness adaptation make SLM agents competitive with frontier-LLM agents?}
    We compare SLM agents with generic and adapted harnesses against frontier-LLM agents on accuracy, cost, and latency.

    \item \textbf{RQ2: How does task diversity impact harness adaptation effectiveness?}
    We examine whether adaptation is more effective for more repetitive tasks.
    
    \item \textbf{RQ3: How does model capability impact harness adaptation effectiveness?}
    We analyze how adaptation effectiveness varies across SLMs with different capabilities.
    
    \item \textbf{RQ4: What harness adaptation strategies are most common for improving SLM performance?}
    We inspect optimized harnesses through our framework (Section~\ref{sec:theory}), connecting observed improvements to failure modes and adaptation strategies.
\end{itemize}

\subsection{Experiment Setups}
\label{sec:setup}

We evaluate harness adaptation in a controlled setting where each task is paired with an initial generic harness, several candidate SLMs, and a frontier-LLM baseline. 
For each task-SLM pair, the optimizer searches for specialized harness that maximizes validation scores, and we evaluate the selected harness on a held-out test set. 
This design lets us analyze not only whether adaptation improves performance (RQ1), but also how the gains relate to task properties (RQ2), model capabilities (RQ3), and the types of adaptation strategies employed (RQ4).

\textbf{Tasks.}
We purposefully select seven agentic tasks designed to reflect real-world business deployment scenarios.
These tasks cover diverse routine business workflows such as auditing records, managing budgets, maintaining websites, writing software tests, and refactoring code.
For each task, we curate a dataset of independent instances from public benchmarks.
Each task comes with an objective evaluation metric, such as exact-match decisions, executable checks, or runnable code.

To study how task diversity affects the effectiveness of harness adaptation (RQ2),
we curate these tasks with varying diversity:
Lower-diversity tasks consist of instances that differ mainly in local inputs but follow the same procedure, whereas higher-diversity tasks include instances that require distinct solution strategies or control flows.

For each task, we use a 20/20/60 train/validation/test split.
The meta-agent uses the training set to collect trajectories and propose harness updates, while the validation set is used to evaluate and select candidate harnesses.
The test set is held out for the final evaluation of the selected harness.
The tasks are summarized in Table~\ref{tab:task-overview}.

\definecolor{rose}{rgb}{1, 0.89, 0.88}\definecolor{lightgreen}{HTML}{e5f5e0}
\definecolor{darkgreen}{HTML}{a1d99b}

\newcommand{\markpoor}{\cellcolor{rose}} %
\newcommand{\markgood}{\cellcolor{lightgreen}}
\newcommand{\markbest}{\cellcolor{darkgreen}}

\newlength{\scorewidth}
\setlength{\scorewidth}{2.3em}

\newcommand{\scorebox}[2]{%
  \begingroup
  \setlength{\fboxsep}{1pt}
  \makebox[\scorewidth][r]{%
    \colorbox{#1}{\rule{0pt}{1.45ex}\makebox[2.2em][r]{#2}}%
  }%
  \endgroup
}

\newcommand{\cell}[3]{%
  \begin{tabular}[t]{@{}r@{}}
    {#1} \\
    {\footnotesize\textcolor{gray}{\makebox[\scorewidth][r]{#2}}} \\
    {\footnotesize\textcolor{gray}{\makebox[\scorewidth][r]{#3}}}
  \end{tabular}%
}

\newcommand{\cellpoor}[3]{%
    \markpoor
  \begin{tabular}[t]{@{}r@{}}
    \scorebox{rose}{#1} \\
    {\footnotesize\textcolor{gray}{\makebox[\scorewidth][r]{#2}}} \\
    {\footnotesize\textcolor{gray}{\makebox[\scorewidth][r]{#3}}}
  \end{tabular}%
}

\newcommand{\cellgood}[3]{%
    \markgood
  \begin{tabular}[t]{@{}r@{}}
    \scorebox{lightgreen}{#1} \\
    {\footnotesize\textcolor{gray}{\makebox[\scorewidth][r]{#2}}} \\
    {\footnotesize\textcolor{gray}{\makebox[\scorewidth][r]{#3}}}
  \end{tabular}%
}

\newcommand{\cellbest}[3]{%
    \markbest
  \begin{tabular}[t]{@{}r@{}}
    \scorebox{darkgreen}{#1} \\
    {\footnotesize\textcolor{gray}{\makebox[\scorewidth][r]{#2}}} \\
    {\footnotesize\textcolor{gray}{\makebox[\scorewidth][r]{#3}}}
  \end{tabular}%
}

\begin{table*}[t]
  \centering
  \caption{Summary of performance and inference cost across evaluation settings. Each cell reports accuracy (top), average cost per instance in USD (middle), and average end-to-end latency (bottom).}
  \label{tab:main-results}
  \small
  \setlength{\tabcolsep}{4.5pt}
  \begin{tabular}{@{}>{\raggedright\arraybackslash}p{4.25cm}rrrrrrrr@{}}
  \toprule
  \textbf{Model} & \textbf{Attn.} & \textbf{Budget} & \textbf{Stock} & \textbf{Anom.} & \textbf{Playwr.} & \textbf{Web.} & \textbf{Refact.} & \textbf{Avg.} \\
  \midrule
  \multicolumn{9}{@{}l@{}}{\footnotesize\textit{\textcolor{adaptgray}{Large Language Models}}} \\

  \texttt{gemini-3.1-pro-preview}
    & \cell{96.5}{\$0.894}{206s}
    & \cell{97.3}{\$0.219}{70s}
    & \cell{86.7}{\$5.785}{319s}
    & \cell{98.9}{\$1.450}{148s}
    & \cell{85.7}{\$0.667}{213s}
    & \cell{76.7}{\$2.011}{157s}
    & \cell{86.1}{\$1.116}{155s}
    & \cell{89.7}{\$1.735}{181s} \\
  \midrule

  \multicolumn{9}{@{}l@{}}{\footnotesize\textit{\textcolor{adaptgray}{Small Language Models}}} \\

  \texttt{gemma-4-26b-a4b}
    & \cell{91.7}{\$0.025}{116s}
    & \cell{75.0}{\$0.039}{77s}
    & \cell{5.0}{\$0.041}{428s}
    & \cell{5.0}{\$0.058}{367s}
    & \cell{9.9}{\$0.018}{461s}
    & \cell{1.1}{\$0.071}{425s}
    & \cell{32.2}{\$0.046}{423s}
    & \cell{31.4}{\$0.043}{328s} \\
    \addlinespace

  {\textit{{+ optimized harness}}}
    & \cellbest{95.7}{\$0.027}{215s}
    & \cellbest{98.3}{\$0.017}{33s}
    & \cellgood{58.9}{\$0.089}{177s}
    & \cellbest{99.4}{\$0.033}{61s}
    & \cellbest{98.1}{\$0.021}{85s}
    & \cellgood{45.6}{\$0.165}{206s}
    & \cellgood{65.0}{\$0.147}{171s}
    & \cell{80.2}{\$0.071}{135s} \\
  \midrule

  \texttt{qwen3-coder-30b-a3b}
    & \cell{23.7}{\$0.030}{64s}
    & \cell{61.0}{\$0.025}{73s}
    & \cell{9.4}{\$0.210}{148s}
    & \cell{1.1}{\$0.076}{100s}
    & \cell{31.4}{\$0.058}{176s}
    & \cell{1.1}{\$0.110}{98s}
    & \cell{60.6}{\$0.089}{91s}
    & \cell{26.9}{\$0.085}{107s} \\
    \addlinespace

  {\textit{{+ optimized harness}}}
    & \cellgood{73.7}{\$0.025}{57s}
    & \cellgood{79.3}{\$0.024}{53s}
    & \cellbest{96.7}{\$0.061}{51s}
    & \cellbest{93.9}{\$0.031}{39s}
    & \cellbest{93.6}{\$0.062}{136s}
    & \cellpoor{24.4}{\$0.122}{96s}
    & \cellgood{62.2}{\$0.121}{100s}
    & \cell{74.8}{\$0.064}{76s} \\
  \midrule

  \texttt{ministral-3-8b}
    & \cell{0.2}{\$0.154}{310s}
    & \cell{28.8}{\$0.012}{22s}
    & \cell{0.0}{\$0.099}{132s}
    & \cell{0.0}{\$0.167}{304s}
    & \cell{0.4}{\$0.172}{370s}
    & \cell{5.6}{\$0.056}{65s}
    & \cell{31.7}{\$0.112}{154s}
    & \cell{9.5}{\$0.110}{194s} \\
    \addlinespace

  {\textit{{+ optimized harness}}}
    & \cellgood{63.3}{\$0.002}{9s}
    & \cellgood{53.7}{\$0.037}{95s}
    & \cellpoor{0.0}{\$0.098}{123s}
    & \cellpoor{0.0}{\$0.167}{296s}
    & \cellpoor{22.3}{\$0.181}{401s}
    & \cellpoor{5.6}{\$0.088}{110s}
    & \cellpoor{30.0}{\$0.120}{168s}
    & \cell{25.0}{\$0.099}{172s} \\

  \bottomrule
    \multicolumn{9}{l}{\fcolorbox{black}{darkgreen}{} best \quad \fcolorbox{black}{lightgreen}{} good \quad \fcolorbox{black}{rose}{} poor} 
\end{tabular}
    \vspace{-10pt}

\end{table*}

\textbf{Models.}
We select and evaluate three models that represent the current generation of open-weight SLMs with strong instruction-following and tool-use abilities: \texttt{qwen3-coder-30b-a3b}, \texttt{ministral3-8b}, and \texttt{gemma\-4-26b-a4b}.
We compare them against a frontier LLM \texttt{gemini-3.1-pro-preview} with a generic harness.

\subsection{Methodology}
We organize the experiments around the four research questions introduced above.

\textbf{RQ1: Understanding whether harness adaptation closes the SLM-LLM gap.}
To answer RQ1, we compare three agent configurations on each task: SLMs with the generic harness, the same SLMs with optimized harnesses, and the frontier LLM with the generic harness. 
We measure test accuracy, cost, and latency of each task-model configuration.
Each configuration is run three times, and we report the average result to account for variations across agent runs.
The cost is calculated based on tracked token usages and the reported official API pricing for each model. For \texttt{gemini-3.1-pro-preview} and \texttt{gemma-4-26b-a4b}, we use the API pricing from Google Vertex AI.
For \texttt{qwen3-coder-30b-a3b} and \texttt{ministral3-8b}, we use the API pricing from AWS Bedrock.

For automated harness adaptation, we use \texttt{gemini-3.1\-pro-preview} as the meta-agent model and implement the agent harnesses using \texttt{software-agent-sdk}~\cite{wang2025openhandssoftwareagentsdk}. 
Each optimization run has a budget of \$20,\footnote{We choose a budget of \$20 as we observed optimization saturation in early piloting tasks given this budget. This results in a total budget of \$1260 for our main optimization experiments.}
which covers meta-agent analysis, task-agent rollouts, and candidate harness evaluations.
For each task-model pair, we run the optimization procedure three times and keep the harness with the best validation score.

\textbf{RQ2: Analyzing how task diversity impacts harness adaptation effectiveness.}
Our hypothesis is that higher task diversity lowers adaptation effectiveness by impacting how reusable a harness can be.
If task instances share recurring workflows and failure patterns, a single adapted harness can encode those patterns and improve many instances; otherwise, it is harder to create a comprehensive harness to cover all diverse failures. 
We test the hypothesis with two sets of experiments:

First, we measure task diversity in our task suite and test whether it correlates with optimized performance.
We operationalize task diversity as behavioral variation across LLM agent trajectories. 
For each task, we collect LLM rollouts, extract their tool-call sequences, and compute the average pairwise normalized Levenshtein distance between these sequences. 
This distance serves as a proxy for workflow diversity: tasks whose instances induce similar tool-use sequences have low diversity, whereas tasks whose instances require different tools or tool orderings have higher diversity.

Because this cross-task analysis may be confounded by other task differences, we also construct diversity-controlled variants of two tasks: attendance and budget. 
For each task, we create a pool of templates, where each template represents a distinct workflow. 
In the budget task, for example, templates may require total-cost tracking, remaining-budget calculation, emergency-fund handling, or policy-based approval decisions. 
We then vary the number of templates included in the dataset to control the level of workflow diversity.

\looseness=-1
\textbf{RQ3: Analyzing how model capability impacts harness adaptation effectiveness.}
We hypothesize that higher model capability increases adaptation effectiveness, because it determines how much task difficulty can remain with the SLM.
While harnesses can offload some difficulty, the model must still follow instructions, use tools correctly, and complete the core sub-tasks that cannot be offloaded. 
To analyze the role of model capability, we use the average benchmark scores reported by Artificial Analysis~\cite{artificialanalysis} as a proxy for models' general capability, and use these scores to contextualize performance trends.

\textbf{RQ4: Analyzing how optimized harnesses improve performance.}
To answer RQ4, we manually inspect the optimized harnesses produced for each task-model pair. 
We categorize the discovered edits according to the framework in Section~\ref{sec:theory}, recording which failure modes are addressed and which adaptation strategies are used.

\subsection{Threats to Validity} 
Despite various mitigations described as part of the experimental design above, like all experiments, ours has limitations and should be interpreted accordingly. 
\textbf{Internal validity:} The agent runs and the optimization process are highly stochastic, even with our mitigation of multiple repeated runs. 
\textbf{External validity:} Our findings are based on a particular set of tasks, models, and one optimizer implementation, and may not transfer cleanly to other tasks, future models, or different optimizer implementations.
We focus on tasks with clear success metrics and clean environments, which do not fully capture the complex deployment scenarios of specialized agents.
\textbf{Reproducibility:} Because we use closed-source frontier LLM APIs that are black-box and may evolve over time, exact replication of the optimization may be difficult once these APIs change or get deprecated. 
However, we believe our findings generalize well to any frontier LLMs with strong capabilities.

\subsection{Results}
\label{sec:main_results}

\textbf{SLMs with optimized agent harnesses can recover 89\% of LLM performance at 4\% cost (RQ1)}.
First, we highlight that for 16 out of 21 task-SLM pairs, we have seen a significant performance boost for small language models when their harnesses are optimized, with seven pairs closing the SLM-LLM gap (Table~\ref{tab:main-results}).
The best performing SLM, \texttt{gemma-4-26b-a4b}, can recover 89\% of LLM performance at 4\% cost, with 25\% latency reduction.

While there is a one-time offline optimization cost per task (\$20 in our experiments), 
this cost is amortized after deployment because the optimized harness can be reused across agent runs. 
Based on the per-run cost savings we observe, the optimization cost is already recovered after 13 runs on average across tasks and models.

\begin{figure}[t]
    \centering
    \includegraphics[width=\linewidth]{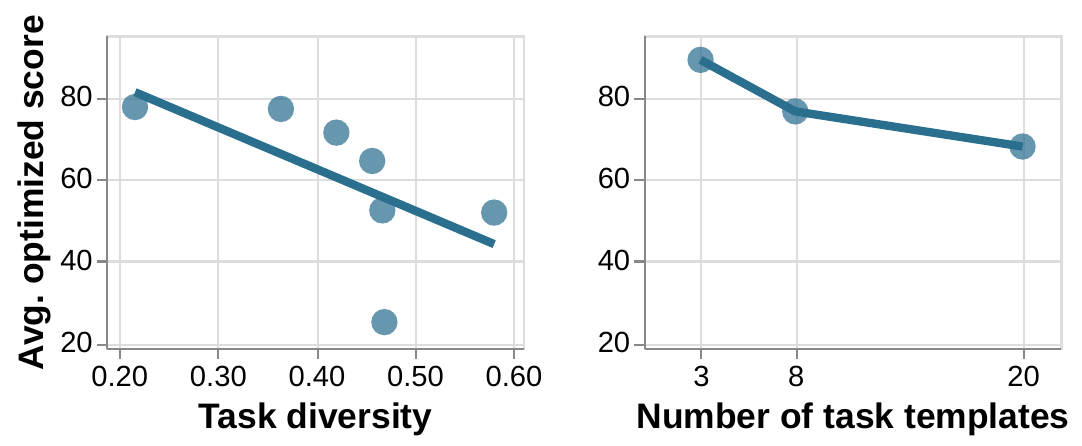}
    \caption{More diverse tasks are harder to optimize. Each point averages across different models (left), or different tasks (right).}
    \label{fig:diversity-impact}
    \vspace{-10pt}
\end{figure}

\textbf{Harness adaptation is more effective on less diverse tasks (RQ2).}
We next analyze when harness adaptation is more effective.
Across the seven tasks, we observed a strong, statistically significant negative correlation (Spearman $\rho = -0.96$) between task diversity and optimized harness performance (Figure~\ref{fig:diversity-impact}, left).
This is further supported by our controlled experiment on task diversity: 
Among the diversity-controlled task variants, we observed that optimized harnesses work worse with more diverse instances, dropping performance from 89.1\% to 68.0\% (Figure~\ref{fig:diversity-impact}, right).

Qualitatively, we observe that tasks like \texttt{attendance\-auditing} are easier to adapt, as their instances have shared failure modes (e.g., missing knowledge and bad output formatting).
In contrast, tasks like \texttt{code-refactoring} are harder to adapt, as different repositories represent different contexts, and different user queries contain divergent requirements for refactoring, which are hard to fully account for in the harness.

\textbf{Harness adaptation is more effective when the model has stronger base capabilities (RQ3).}
Across SLMs, we observe a clear trend that models with stronger capabilities perform better and improve more with optimized harnesses (Figure~\ref{fig:intelligence-impact}).
This reflects how weaker SLMs struggle with the intrinsic task difficulty that cannot be offloaded to the harness.
For example, \texttt{playwright-testing} requires strong {coding} capability, of understanding HTML webpages and writing tests, that cannot be fully delegated to the harness.
Similarly, \texttt{stock-alert} tasks require strong \dimtag{long-context} capabilities, which are difficult to fully capture within a harness.

\textbf{Instruction-following and knowledge failures are the dominant failure modes (RQ4).}
Reviewing the optimized harnesses, we found that the most commonly addressed failure modes are \dimtag{instruction-following} (81\%)  and \dimtag{knowledge} failures (81\%), followed by \dimtag{tool-use} (62\%) and \dimtag{long-context} failures (33\%).
These failure modes are most commonly addressed by adding contexts (86\%), creating tools (43\%), and managing tools (29\%).
For example, the best \texttt{gemma-4-26b} harness for the \texttt{anomaly-detection} task combines three strategies together: it creates a custom \texttt{query\_mock\_bigquery} tool that addresses long-context issues from the default \texttt{bigquery\_run\_query} MCP tool; it filters the agent's active tool set from 40+ MCP tool down to seven, and externalizes the environment's table-naming convention in the system prompt.

Noticeably, we do not see successful harness adaptation of creating sub-agents. 
We hypothesize that this is an artifact from current SLMs' limited agent management capabilities -- i.e., they struggle with tracking and managing sub-agents' work progress and coordinating among different agents.
Alternatively, this can be seen as limitations of the meta-agent, which does not generate a good harness for easier sub-agent management.

\textbf{Different SLMs require different adaptation strategies (RQ4).}
Across different SLMs, we observe major differences in how these models can be improved.
For example, \texttt{ministral-3-8b} sometimes struggles with using file editing tools,
while \texttt{qwen3-coder-30b-a3b} occasionally fails to emit correct JSON tool calls (but instead emit raw XMLs).
Each model's failure patterns require tailored harness adaptation that does not cleanly transfer.
This indicates that, there is no one-size-fits-all harness, and that harness optimization needs to be re-run each time we migrate to a new SLM.

For example, for the \texttt{playwright-testing} task, the optimized harness for \texttt{gemma-4-26b-a4b} adds contexts to externalize knowledge and enforce instructions, and create hooks to enforce successful pytest runs.
Meanwhile, the optimized harness for \texttt{ministral-3-8b} creates new tools to mitigate the model's file-editing failures.

\section{Discussion}
\subsection{How to optimize agent harnesses effectively?}
\label{sec:optimizer-discussion}
Our experiments show it is promising to automatically optimize agent harnesses for SLMs.
Throughout the optimizer development, we also learned a few lessons about building effective harness optimizers, which we share below.

The main design question of building a harness optimizer is on how to the allocate optimization budgets well to find genuine improvements. 
In our optimizer, the budget can be spent on (1) analyzing failures, or (2) evaluating candidate fixes, with multiple iterations: 
$
B = \sum_{t=1}^{T} \left(b_t^{\mathrm{analyze}} + b_t^{\mathrm{evaluate}}\right).
$
Given a fixed budget, we can choose to (1) analyze deeper, (2) evaluate more extensively, or (3) run more iterations.

\textbf{Lesson 1: Improve analysis quality, with better contexts and intelligence.}
Harness optimization is bottlenecked by diagnosis quality -- spending budget on evaluation or additional iterations is useful only when the meta-agent can propose good harness candidates.
We found that the meta-agent performs better when it receives high-fidelity failure evidence rather than aggressively summarized traces.
In early iterations, we converted task trajectories from structured JSON into post-processed markdown, but this sometimes removed details that were useful for understanding failures.
Passing raw JSON worked better in practice: although it is less human-friendly, the frontier meta-agent could manipulate and inspect it effectively while benefiting from the extra information.

The same lesson applies to the choice of meta-agent model.
Replacing \texttt{gemini-3.1-pro-preview} with a cheaper model \texttt{gemini-3-flash-preview} affords more iterations, but the lower-quality diagnoses and edits more than offset the extra exploration, yielding worse final harnesses. 

At the same time, a good frontier model also does not require extensive hand-holding.
We found that showing successful trajectories from frontier models or manually written heuristics about which adaptations match which failure modes did not consistently improve outcomes.
This suggests that, once given faithful failure evidence and an editable harness, a strong meta-agent is well-positioned to infer targeted repairs on its own.

\begin{figure}[t]
    \centering
    \includegraphics[width=\linewidth]{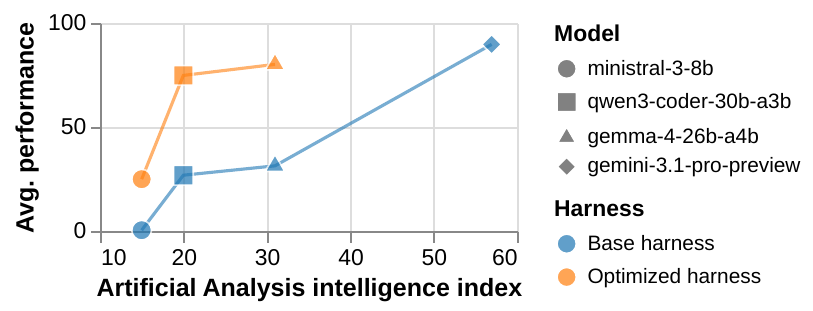}
    \caption{Less capable models are harder to optimize.}
    \label{fig:intelligence-impact}
    \vspace{-5pt}
\end{figure}

\textbf{Lesson 2: Prioritize exploration quality and diversity, not quantity.}
Because full agent evaluation is expensive, harness optimization offers far fewer search steps than single-turn prompt optimization.
This makes the quality and diversity of proposals more important than simply maximizing the number of iterations.
For quality, it is important to use a good frontier model for proposing new harnesses, as we have discussed before.
For diversity, we observed two helpful design choices:

First, keep a memory of prior attempts. Without logging previous edits and their outcomes, the meta-agent tends to rediscover similar fixes, wasting budget on redundant local exploration.
Second, run several independent searches than to spend the same budget on one long trajectory.
Multiple runs cover more distinct regions of the design space and increase the chance that at least one run discovers a strong adaptation.

\subsection{Implications}

\textbf{For agent developers.}
When developing new agents for routine business tasks where cost matters, developers should consider and experiment with a wide range of SLMs, instead of throwing all tasks to a frontier LLM.
SLMs can be particularly effective, when the task is more repetitive and more task demands can be offloaded to the harness.
To choose an appropriate SLM, a good proxy is benchmark scores (cf. Figure~\ref{fig:intelligence-impact}), which we observe to correlate with better performance after harness adaptations.
Heuristically, small Mixture-of-Experts (MoE) models (e.g., \texttt{gemma-4-26b-a4b}) are good candidates, as they provide great model capabilities with a low inference cost.

To adapt SLM harnesses, developers should consider an automated approach beyond manual engineering. 
This would require them to curate task instances, with clear evaluation criteria, to optimize against.
Generally, we observe a meta-agent powered by frontier LLMs to be fairly strong on the harness optimization task.

\textbf{For model developers.}
Training practically useful SLMs goes beyond optimizing benchmark scores.
To achieve better cost-performance tradeoffs, model developers should \textit{optimize for harness compatibility}.
For example, a good SLM should reliably work with various prompt scaffolds (\dimtag{instruction-following}) and tool scaffolds (\dimtag{tool-use}).
This would require SLMs to be exposed to a diverse range of environments, with different prompts and tools available, during their training.

\looseness=-1
\textbf{For researchers.}
Automated harness optimization is an emerging research topic with many open challenges and opportunities.
Designing better optimizers (cf. Section~\ref{sec:optimizer-discussion}) will benefit from evidence from larger-scale experiments and a more thorough exploration of the design space.
For example, future research can develop more fine-grained task and model profiles based on individual failure modes / capabilities to better predict when a task is likely to benefit from harness adaptation, and what kinds of harness edits are most likely to help.

Another promising direction is to move beyond a single adapted harness per task.
For more diverse tasks that contain clusters of repetitive instances, researchers could study \textit{mixtures-of-harnesses}: collections of specialized harnesses paired with routing systems that select the appropriate harness for each instance.
This may avoid forcing one harness to cover too many cases, which can make the harness overly complex and difficult for SLMs to work with.
Similarly, when task distributions drift over time, online monitoring and adaptation systems could maintain multiple harness versions and update routing or harness as new failures emerge.

\section{Related Work}
\label{sec:related}

Building effective agent harnesses is a widely discussed topic among practitioners~\cite{anthropic2024building_agents}, and we have outlined the major capabilities and harness components that the literature discusses in Section~\ref{sec:theory}.
Here, we discuss additional related work on using SLMs for agentic tasks and automating agent harness design.

\textbf{Using Small Language Models for Agentic Tasks.}
With growing interest in using SLMs for agentic tasks~\cite{belcak2025small},
many industry labs have been training SLMs with strong agentic capabilities~\cite{qwen35blog}.
These provide us with a good foundation to start with -- without basic \dimtag{tool-use}, or \dimtag{instruction-following} capabilities, it would be very hard-pressed to build any effective agents to mitigate the model's limitations.

On the harness side, we have seen work emerging to address SLM limitations.
For example, Lee et al.~\cite{lee2025don} propose to adapt tool schema to help with smaller models with limited \dimtag{tool-use} capabilities with unfamiliar schema.
Srivastava et al.~\cite{srivastava2026effgen} design a whole set of human-engineered heuristics to mitigate SLM limitations (e.g., \dimtag{instruction-following} with long prompts).
These approaches heavily depend on observations of specific SLM behaviors and may not generalize well to newer models with improved capabilities.
In contrast, our work intentionally does not make assumptions on what SLMs are capable of, but employs a meta-agent to empirically diagnose their trajectories and create tailored harnesses.

\textbf{Automating Agent Design.}
Automating agent design has been explored a lot in the literature.
ADAS~\cite{hu2024automated} is one of the early work that uses an LLM to optimize agents defined in code.
Other work has extended this to agents with explicitly modular components (e.g., planning, memory)~\cite{shang2024agentsquare}, to topology in multi-agent systems~\cite{zhang2025multi, zhou2025multi}, etc.
Most of these work focus on agentic systems with pre-defined workflows (vs. tool-use agents in our work) and tasks that do not require interacting with environments.
They also tend to focus on creating harnesses that improve performance by increasing model compute (e.g., test-time scaling), instead of by moving work from models to harnesses as in our work.

Another line of work of self-improving agents~\cite{zhang2025darwin, xia2025live} have explored using agents to modify its own code.
A common observation is that these agent often make new tools to improve task performance.
Similarly, Lee et al.~\cite{lee2026meta} have explored automating agent harness designs with a meta agent.
These work's goals are to improve agent performance on given tasks, often the ones that are challenging even for frontier models (e.g., TerminalBench~\cite{merrill2026terminal}),
while we focus on identifying SLM harnesses that recover frontier-agent performance at lower cost.
Concurrent work on Life-Harness~\cite{xu2026adapting} also adapts agent harnesses for frozen models, but focuses on evolving a harness that transfers across benchmark environments and model backbones, whereas we study harness adaptation as a cost-performance strategy for making specialized SLM agents competitive with frontier-LLM agents on routine business tasks.

\section{Conclusion}
\label{sec:conclusion}
In this work, we present a first in-depth empirical analysis of if, when, and how harness adaptations can afford construction of (cost-)effective SLM agents.
We show that, given a well-described design space, a harness optimizer can automatically identify effective harness adaptations that significantly boost SLM performance, making it possible to build effective agents at a fraction of the cost of LLMs.
Our analysis found that this works best for more repetitive tasks and SLM with strong base capabilities.
We can also explain \textit{why} the optimized harnesses are effective, by connecting the changes to the observed failure modes and established adaptation strategies.

\section*{Acknowledgment}
This work is supported by Apple gift funding, the Amazon AI Research gift fund, and the Gemma Academic Program GCP Credit Award.
We thank Andrew Walkingshaw and Eric Liang Yang for their discussion and feedback throughout the project.
We also thank Manisha Mukherjee and Vasu Vikram for their feedback on this work.

\bibliographystyle{IEEEtran}
\bibliography{main}

\end{document}